# Thermo-optical pulsing in a microresonator filtered fiber-laser: a route towards all-optical control and synchronization

MAXWELL ROWLEY,[1] BENJAMIN WETZEL,[1,2,*] LUIGI DI LAURO,[1] JUAN S. TOTERO GONGORA,[1] HUALONG BAO,[1] JONATHAN SILVER,[3] LEONARDO DEL BINO,[3] PASCAL DEL' HAYE,[3] MARCO PECCIANTI,[1] ALESSIA PASQUAZI,[1,**]

[1]*Emergent Photonics Lab (EPic), Dept. of Physics and Astronomy, University of Sussex, Brighton, BN1 9QH, United Kingdom*
[2]*Xlim Research Institute, CNRS UMR 7252, Université de Limoges, 87060 Limoges, France*
[3]*National Physical Laboratory (NPL), Teddington TW11 0LW, United Kingdom*
*\*b.wetzel@sussex.ac.uk*
*\*\*a.pasquazi@sussex.ac.uk*

**Abstract**: We report on 'slow' pulsing dynamics in a silica resonator-based laser system: by nesting a high-Q rod-resonator inside an amplifying fiber cavity, we demonstrate that trains of microsecond pulses can be generated with repetition rates in the hundreds of kilohertz. We show that such pulses are produced with a period equivalent to several hundreds of laser cavity roundtrips via the interaction between the gain dynamics in the fiber cavity and the thermo-optical effects in the high-Q resonator. Experiments reveal that the pulsing properties can be controlled by adjusting the amplifying fiber cavity parameters. Our results, confirmed by numerical simulations, provide useful insights on the dynamical onset of complex self-organization phenomena in resonator-based laser systems where thermo-optical effects play an active role. In addition, we show how the thermal state of the resonator can be probed and even modified by an external, counter-propagating optical field, thus hinting towards novel approaches for all-optical control and sensing applications.

## 1. Introduction

Within the field of nonlinear optics, the long cavity lifetime and reduced mode-area provided by high-Q whispering gallery mode (WGM) resonators has been demonstrated as an efficient way to achieve important Kerr nonlinearity enhancement, and thus unlock numerous functionalities at reasonable power levels compatible with continuous-wave (CW) operation [1]. Besides the rather conventional excitation of microresonators using an external optical field [2–4], there have been significant efforts over the last few years to utilize micro-cavities for the development of novel laser configurations [5], especially with the prospect of system miniaturization. Applications of these embedded resonator systems include compact frequency comb generation [6–11], including self-injection locking schemes [12–15], passively mode-locked nanosecond pulsing [16], microdisk lasers development [17,18] or even the generation of quantum frequency combs [19].

Such applications are possible due to strong and desirable nonlinear effects in these devices (e.g. Kerr nonlinearity), which can be effectively paired with more complex dynamics. Indeed, in these resonant structures, the strong field enhancement is usually associated with additional intensity-dependent effects which can be considered detrimental: free carrier absorption in silicon [20], Raman and Brillouin scattering [21], as well as thermo-optical [22] and opto-mechanical [23] effects can play a disruptive role in numerous dynamical processes.

Of particular interest here is the temperature-dependent modification of the optical refractive index and/or mode volume, ubiquitous in all WGM resonators. Understanding and controlling this thermally-induced frequency drift has been key in the generation of cavity solitons by externally seeding WGM cavities with a CW field [7]. Indeed, such schemes typically require a form of active monitoring and feedback control to ensure stable operation. There are however examples of the beneficial use of these adverse nonlinear effects, notably for the demonstration of optomechanical chaos transfer [24], the generation of giant pulses [25], thermal sensing [26], regenerative pulsation [27] and self-sustained pulsation [28–30], also controlled by parametric nonlinearity [31].

Interestingly, using an external amplifying cavity structure to passively stabilize slow temperature drifts and circumvent thermo-optical effects was already proposed and demonstrated in microresonators [6]. Yet, to date, the dynamical interactions between these two non-instantaneous physical processes still remain widely unexplored beyond their stabilization capabilities.

Here we report on a thermo-optically driven pulsing mechanism demonstrated in a laser system consisting of a fused-silica WGM resonators [32] directly nested in an external fiber amplifying loop. In our system, the complex dynamical interaction between the thermo-optical effect in the resonator and the slow response of the gain medium leads to a sustained self-pulsing. Such dynamical behavior, typically encountered in nonlinear optical cavities exhibiting a nonlinear relaxation time significantly longer than the cavity round trip time [33], is here observed experimentally. We reveal that the observed pulse train characteristics can be directly controlled by adjusting the laser cavity parameters. Additionally, we show a thermo-optically mediated transfer of the dynamics of this lasing mode to an externally coupled probe field in the microresonator. The experimental results, exhibiting a variety of self-organization effects and multistable dynamics, are supported by numerical simulations exhibiting good qualitative agreement. Our results pave the way towards non-instantaneous all-optical control of the resonator dynamics along with potentially wavelength- and/or spatial-mode-independent signal processing and synchronization features.

## 2. Experimental setup

Our experimental setup is shown in Fig. 1(a): we nest a high-Q (>$10^6$) fused-silica rod resonator coupled via tapered fibers (see inset) inside an Erbium-doped fiber amplifier (EDFA) loop cavity. An isolator is incorporated within the fiber cavity to ensure unidirectional operation (see blue arrows). The overall intracavity power can be adjusted by the EDFA pump current, as well as additional losses induced by a variable optical attenuator. The 4-port resonator transmission properties can additionally be probed via a continuous wave (CW) laser with tunable wavelength (Tunics-plus), counterpropagating within the resonator. Both the intracavity field (blue) and optional counterpropagating CW probe properties can be measured temporally using a fast oscilloscope and photodiodes (200 MHz oscilloscope bandwidth) and spectrally using an optical spectrum analyzer (Anritsu MS9740A with 70 pm resolution). The intracavity field is probed using a 10% output fiber coupler (90:10) and eventual back-reflections between the frequency-detuned counterpropagating fields are avoided by the presence of isolators and a wavelength division multiplexer.

The broadband EDFA spectrum transmitted by the resonator is shown in Fig. 1(b), and further filtered by a bandpass filter, consisting of a fiber Bragg grating and a circulator. The bandpass filter is centered at 1545.064 nm and exhibits a 33.7 GHz bandwidth (see blue shading). As illustrated in Fig.1(c), this bandwidth is slightly larger than the resonator free spectral range (FSR) of 24.1 GHz. The fine transmission spectrum reveals a structure featured with multiple spatial modes. However, as seen in Figs. 1(d-e), a careful selection of the resonator input polarization allows for the selection of only one (or few) predominant resonances to oscillate in the cavity within the bandpass filter bandwidth, and thus minimizes mode competition (see Fig. 1(f)). In the main cavity, all components are polarization-maintaining besides two fiber polarization controllers before and after the tapers coupled to the resonator, to respectively select the main coupled resonance and minimize the losses between the resonator and polarization-maintaining cavity components. Fig. 1(g) shows the selected resonance (purple) and the Lorentzian fit (dashed black) obtained by accounting for the 400 kHz linewidth of our CW sweeping laser. We found a coupled linewidth (FWHM) of 27.2 MHz thus yielding a 7.1 million loaded Q-factor at 1545 nm (the intrinsic Q-factor is around $10^8$) and encompassing a single mode of the 6.7 m main amplifying fiber cavity (i.e. 30.9 MHz FSR).

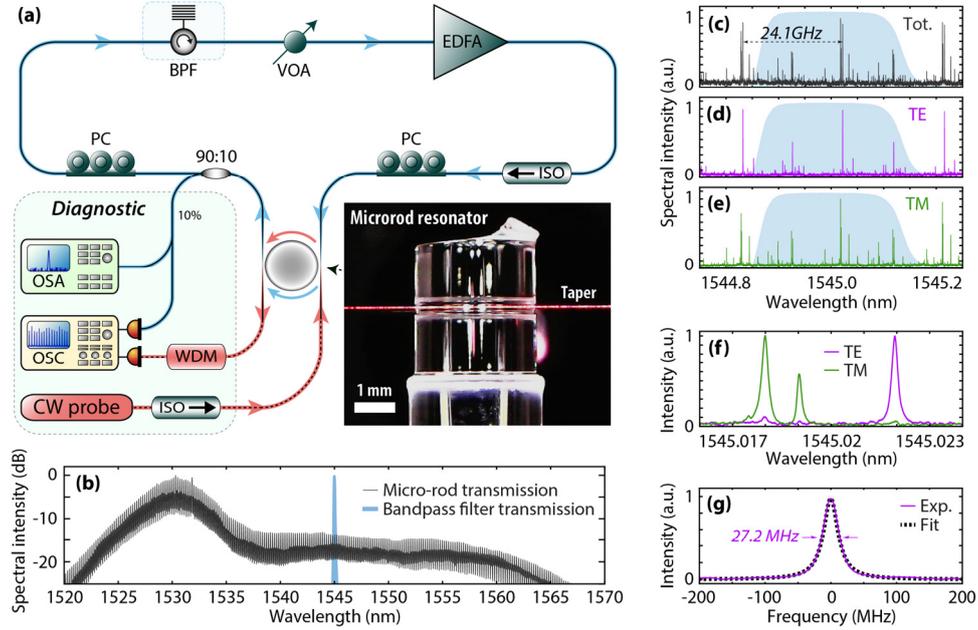

Fig. 1. (a) Schematic of the experimental setup. PC, fiber polarization controller; VOA, variable optical attenuator; BPF, bandpass filter; EDFA, Erbium-doped fiber amplifier; ISO, optical isolator; WDM, wavelength-division multiplexer; OSC, oscilloscope; OSA, optical spectrum analyzer. Inset: Side view of the silica rod resonator and coupled fiber used for optical injection. (b) OSA transmission spectra of the resonator (grey) and BPF (blue shading) when seeded by the EDFA. (c-e) High-resolution transmission spectra are obtained by CW wavelength sweep using either unpolarized (c) or polarized transverse electric (TE) (d) and transverse magnetic (TM) (e) light. (f) Zoom on the predominant spatial modes oscillating for each polarization. (g) Spectrum of the TE main resonance and Lorentzian fit (dashed black line).

## 3. Results and Discussion

When the system operation is suitably adjusted, stable and sustained self-pulsing is observed, as illustrated in Figure 2(a). In this lasing regime, the average pulsing period is typically 15 μs with a pulse duration on the order of a microsecond (see below for details). Here, the pulsing operation is rather stable with a peak intensity relative noise below 5% RMS deviation. Figure 2(b) shows a slight pulse asymmetry, where an elongation is observed in the trailing edge, a feature typical of dynamics associated with long timescales relative to the cavity lifetime. The corresponding spectrum, shown in Fig. 2(c), features a single spectral line with a linewidth in agreement with a single mode resonance oscillation (when considering the limited OSA spectral resolution) while the phase space portrait, represented in Fig. 2(d), illustrates the presence of an attractor associated with the self-pulsing behavior observed in this system.

In fact, such self-pulsing dynamics are notorious in many complex and nonlinear systems, and are typically encountered in optical lasing architectures exhibiting bistability, where the relaxation time of the nonlinear effect is significantly longer than the cavity round trip time [29,30,33]. In our case, the pulse train period (~ 15 μs) corresponds to over 450 cavity roundtrips but is, conversely, in the order of magnitude of the dynamical timescales at play within our system, namely the thermo-optical effects in the resonator and gain recovery in the main fiber cavity.

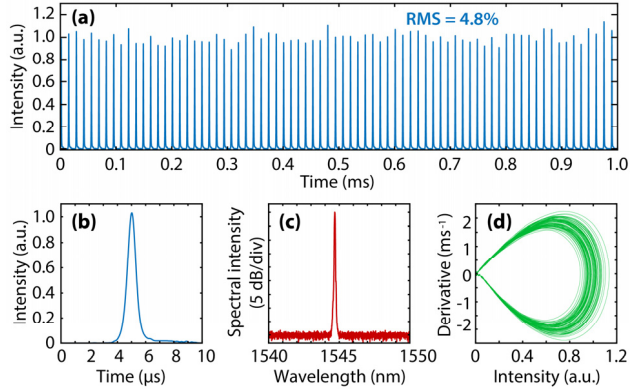

Fig. 2. (a) Typical pulse trace measured experimentally using the setup shown in Fig. 1. (b) Zoom on a single pulse period. (c) Corresponding OSA spectrum. (d) Phase space portrait of the pulse train.

In order to obtain insight into the dynamics observed experimentally, we employ a numerical model based on coupled-mode theory [34]. In particular, we follow the field evolution in both the resonator and fiber cavities, assuming an intensity-dependent frequency shift induced thermally in the resonator. This thermo-optical behavior is schematically illustrated in Fig. 3(a) and the overall evolution of the system is expressed as:

$$\begin{cases} \partial_t a = -\pi \Delta F_A a + i|a|^2 a + i\Delta\omega_T a + \sqrt{\theta}\dfrac{b}{\sqrt{T_A T_B}} & (1) \\ \partial_t b = -i\Delta\omega b + \left(g - \alpha - \dfrac{1}{T_B}\right) b + \sqrt{\theta}\dfrac{a}{\sqrt{T_A T_B}} & (2) \\ \partial_t g = \dfrac{(g_0 - g)}{T_g} - R_g |b|^2 g & (3) \\ \partial_t \Delta\omega_T = -\dfrac{1}{T_T}(\Delta\omega_T - R_T |a|^2) & (4) \end{cases}$$

Here, $a$ and $b$ are the normalized mode amplitudes in the resonator and main cavity, respectively [35, 36]. We model the EDFA in terms of a standard set of Maxwell-Bloch (MB) equations for a two-level system [37], so that $g$ corresponds to the gain of the main cavity observed with an estimated population relaxion time $T_g$ = 900 µs [38]. The thermo-optical frequency shift of the resonator $\Delta\omega_T = 2\pi\Delta\nu$ is intensity-dependent with an estimated relaxation time constant $T_T$ = 3 µs [22]. The resonator and main cavity round-trip times used in our simulations correspond to the experimental values and respectively $T_A = 1/F_A$ = 41.5 ps and $T_B = 1/F_B$ = 29.8 ns. Similarly, we assume a resonator with a linewidth of $\Delta F_A$ = 27.2 MHz, which is strongly dominated by the coupling to the two tapered fibers, so that the coupling constant between the main cavity and the resonator can be estimated as $\theta = \pi \Delta F_A T_A$. It is worth mentioning that, in our equations, the mode amplitudes are normalized with respect to the Kerr nonlinear mode coefficient $\Gamma_K = 4.82 \times 10^{15}$ J$^{-1}$s$^{-1}$ estimated in resonator. Specifically, the normalized amplitude in the resonator $a$ can be derived from the dimensional mode energy $A$ (in Joules) so that $|a|^2 = \Gamma_K A$. The main cavity mode follows the same normalization where $|b|^2 = \Gamma_K B$ and the strength of the intensity-dependent thermal nonlinearity $\Gamma_T$ is estimated to be approximately two orders of magnitude larger than the instantaneous Kerr effect in the resonator, thus yielding $R_T = \Gamma_T/\Gamma_K$ = 100 [39]. The gain-main cavity field coupling constant, corresponding to the inverse of the saturation power, is given as $R_g = \eta_g/\Gamma_K$ = 69.6 where we estimated a gain coefficient $\eta_g = 3.35 \times 10^{17}$ J$^{-1}$s$^{-1}$ assuming a fast decay-time of 400 fs for the EDFA doping concentration used in our experiment [35]. In our simulations, we can observe a wide variety of dynamics depending on the static cavity parameters, namely the main cavity losses $\alpha$ and initial gain $g_0$, as well as the the cold cavity angular frequency mismatch between the central resonance and the main-cavity mode $\Delta\omega$. However, for selected parameters, numerical simulations of the system display sustained self-pulsing in good agreement with the dynamical behaviour observed in our experiments.

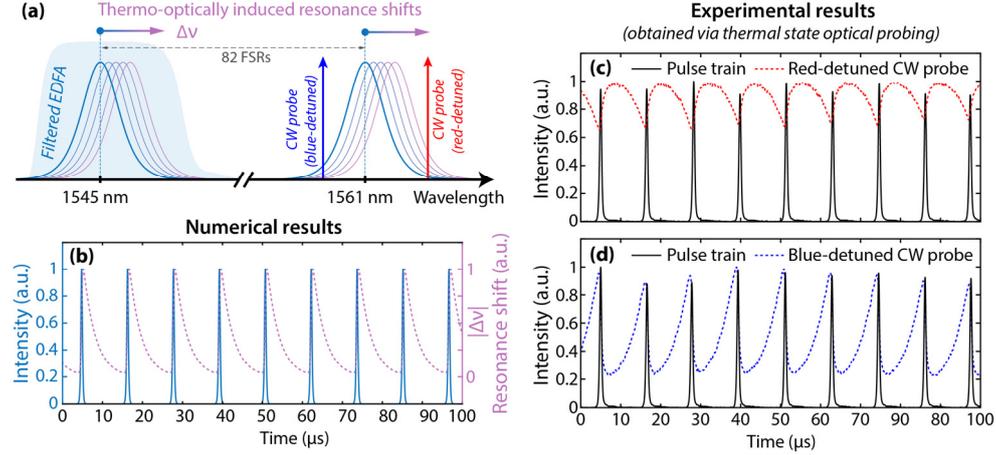

Fig. 3. (a) Schematic of the physical process leading to self-pulsing via thermally-induced resonance shifting in our setup: The loaded Lorentzian resonator mode within the filtered bandwidth of the EDFA (shaded blue region around 1545 nm) dynamically shifts by an amount $\Delta v$ via the thermo-optical effect (see gradient arrow and colored shadings). This shift is also expected across all resonances and as such, can be readily measured experimentally using a CW probe weakly coupled and slightly detuned from a resonance far outside the main cavity gain bandwidth (see e.g. red and blue arrows around 1561 nm). (b) Numerical results obtained using the coupled mode equations of Eqs. (1-4) for parameters yielding self-pulsing behavior. The field intensity in the resonator, $|a|^2$ (solid blue line – see Eq. 1) shows a train of pulses, associated with a periodic resonance frequency shift $\Delta v$, induced by thermo-optical effects (dashed purple line – see Eq. 4). (c) Example of pulse train obtained experimentally (solid black line) and corresponding transmission of a red-detuned (dashed red line) CW field, used to retrieve the dynamical resonance spectral shift as shown in (a). (d) Same measurements using a blue-detuned CW probe (dashed blue line).

The numerical results are illustrated in Fig. 3(b), where the formation of pulses (solid blue line) is associated with a rapid frequency shift $\Delta v$ of the seeded resonance (dashed purple line), followed by a slower recovery of the initial (blue-detuned) resonance frequency during the absence of pulse emission. Physically, the dynamical mechanisms responsible for such operation are qualitatively described in Figure 3(a): the length of the main cavity, combined with the band-pass filter (see Fig 1(b)), ensure that only a predominant cavity mode lies within the resonator linewidth $\Delta F_A$. The oscillation of such a mode induces a thermo-optical refractive index variation in the resonator, associated with a varying detuning of the main cavity mode within the shifting resonance. This in turn gives rise to a thermo-optically modulated loss mechanism, with a characteristic timescale much longer than the roundtrip time of the laser cavity. The dynamical interaction between the gain recovery mechanisms and the relaxation of the thermo-optically induced losses - both in the microseconds regime - can therefore lead to the sustained self-pulsing behavior observed in Fig. 3(b).

Experimentally, the dynamical frequency shift of the resonator (and associated cavity losses) can be readily measured using a weak (<100 µW) counter-propagating CW probe laser, as illustrated in Figure 1(a). In our case, depending on the spectral detuning of the CW probe weakly coupled to the resonance around 1561 nm, a variation in the transmitted power is expected to be observed. In particular, as illustrated in Fig. 3(a), depending on whether the CW probe is red- or blue-detuned with respect to the resonance, one would respectively assume an increase or decrease of the transmitted power when the resonance is thermo-optically frequency shifted by the pulses emitted within the laser cavity. This behavior has been experimentally measured in our setup, and reported in Fig. 3(c) and (d), respectively. Stable self-pulsing is observed in the cavity (black line) while the CW probe exhibits a synchronous and periodic variation in transmission whose sign and magnitude depend on its spectral position compared to the resonance (dashed red and blue lines). In fact, the measured variation of the CW probe transmission here corresponds to the spectral convolution of the CW linewidth with the dynamically shifting resonance. Assuming a Lorentzian shape with a linewidth of 27 MHz (FWHM), and repeating these measurements for various CW detunings, we estimated that a typical 10 MHz spectral shift was observed for pulses circulating within the external gain cavity with an approximate ~10 mW peak power. This corresponds to a thermo-optical frequency shift of approximately 1 MHz/mW, a value in line with the measurements reported in the literature for the spectral shift observed for fused silica rod resonators [40] and extrapolated to the 24 GHz FSR of the resonator used in our experiments.

In order to gain further insights on the pulse formation mechanism, we perform an analysis of the generated pulse train properties depending on the main cavity static parameters. These results, obtained by varying the transmission (i.e. losses) and gain of the laser cavity both experimentally and numerically, are summarized in Fig. 4. From Fig. 4(a) and Fig. 4(b), respectively obtained by increasing or decreasing the EDFA gain for various transmission conditions, one can first notice multistable effects: in this case, the pulse properties (i.e. durations and peak powers) not only depend on the cavity parameters, but also on the its previous state (i.e. gain increase/decrease). Although various pulsing properties can be achieved for different transmission/gain conditions, we have observed that higher cavity gain resulted in more intense pulses with a shorter duration. This result is in fact not surprising, as higher gain intrinsically implies a reduction of the load time required to induce a sufficient thermo-optical frequency shift of the resonance (responsible for additional losses in the cavity, and thus defining the overall pulse duration). This observation was confirmed experimentally by replacing the previous EDFA (Amonics, with length of 1.2 m and 23 dBm output saturation power) with a longer, higher power amplifier (HP-EDFA – MENLO P250, with length of 5.7 m and a higher saturation power of 27 dBm). In this case, the overall cavity roundtrip was significantly longer but with similar losses. Yet, as seen in the inset of Fig. 4(a), the higher gain accessible with the HP-EDFA further enables the generation of pulses with superior peak powers (and correspondingly reduced durations). Indeed, increasing the system losses also leads to a reduction of the pulse duration, as the thermo-optically induced frequency drift is dissipated faster by the higher losses in the main cavity. Interestingly however, in the experiments reported here, the overall variation in the cavity losses are relatively small (within a single order of magnitude) and intrinsically limited by the gain range of the EDFAs used to observe self-pulsing.

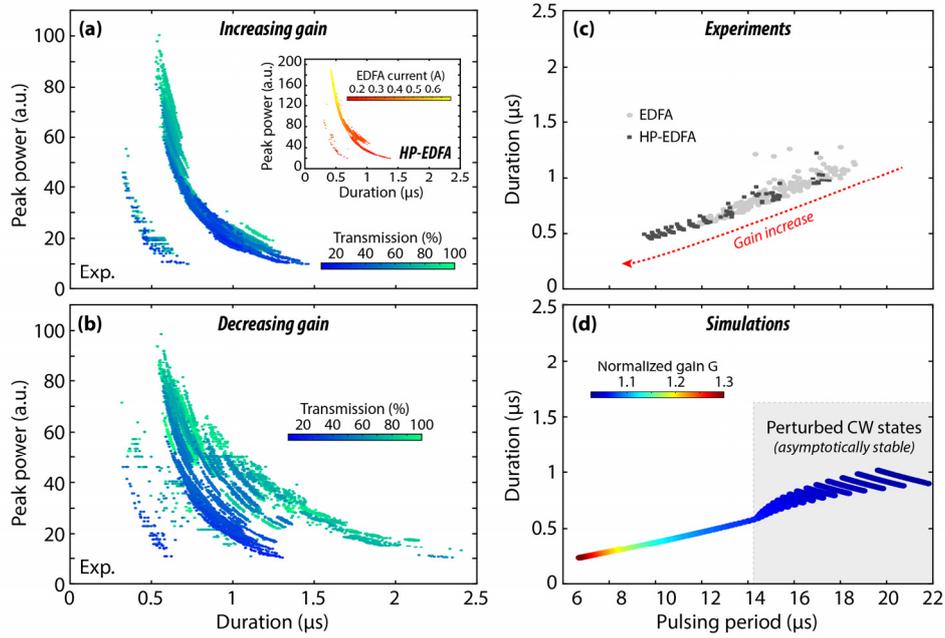

Fig. 4. (a, b) Scatter plots of pulse peak power and duration obtained experimentally using the setup shown in Fig. 1. Here, data are extracted for various cavity transmission values (see color scale) and EDFA gain settings, while either increasing (a) or decreasing (b) the EDFA gain. The inset in (a) corresponds to the same scatter plot, obtained experimentally by replacing the initial EDFA in the setup by a longer and higher power gain medium (HP-EDFA). Here, the data are also obtained for various cavity transmission values while increasing the EDFA gain, but the color scale instead represents the gain medium pumping current. (c) Scatter plot of the mean pulse duration and period obtained experimentally for various loss values and upward tuning of the gain. This analysis was performed using both available amplifiers (see legend) and by averaging the properties extracted from each ms-long experimental trace. (d) Corresponding properties retrieved from numerical simulations using a selected loss parameter α. Scatter points of the pulse properties are displayed for various values of normalized gain $G = g_0/\alpha$ (see color scale). The numerical results are obtained using the coupled-mode equations of Eqs. (1-4) for selected parameters yielding self-pulsing behavior. The pulse properties were extracted from the numerical field intensity via the same post-processing used on the experimental datasets. The grey shading region (featuring clusters of points instead of a single point), corresponds to a dynamical regime where simulations exhibit self-pulsing only sustained for a limited period (i.e. in a slow transient ultimately leading to a stable CW solution).

Within this experimental regime shown in Fig. 4(c), we can observe that the self-pulsing period is indeed directly related to the associated pulse duration, in a manner almost independent of the selected amplifier (and length) but mainly depending on the gain value itself. These results are confirmed by numerical simulations, as reported in Fig. 4(d), where we have observed that for a fixed loss parameter α, the pulse duration and self-pulsing period were indeed correlated, and readily controllable by adjusting the gain $G$ in the cavity. Physically, this behavior can be explained by the relationship between the system parameters on the gain recovery time with respect to the resonator linewidth and its susceptibility to thermo-optical effects. In our numerical simulation, we observe stable self-pulsing behavior for pulse periods between 6 and 14 us, which agrees qualitatively with the experimentally observed range of parameters, spanning from 10 to 19 us. The small discrepancy can be attributed to the absence of higher order effects such as the presence of the fiber nonlinearity in the model and slight discrepancies in the simulation parameters. Interestingly, the numerical simulations point out that the system still shows some self-pulsing behavior also for pulses longer than 14 us. These simulations were characterized by a long-term disappearance of the self-pulsing after thermalization of the system, a condition which cannot be easily met in the experiment due to the presence of noise.

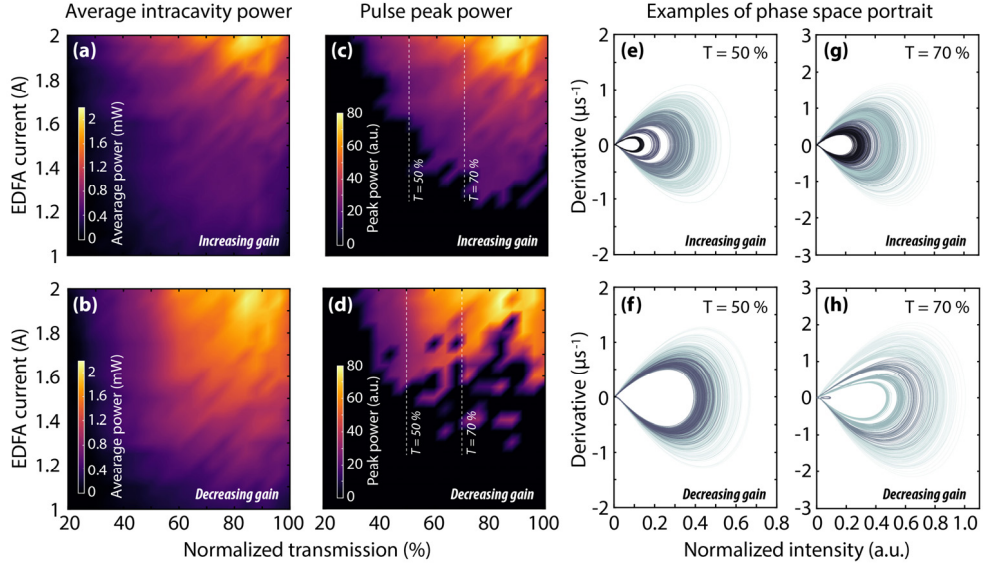

Fig. 5. (a-b) False color maps showing the average intracavity power (see color scale) for experimental measurements of the laser cavity operation. Measurements are performed for a range of loss values while increasing (a) or decreasing (b) the gain. (c-d) Corresponding map of the pulse peak powers displaying clear regions of self-pulsing operation (i.e. black areas are representative of parameters for which self-pulsing is absent - see color scale). (e-h) Examples of phase space portraits retrieved from experimental data measured at selected transmission of 50% (e, f) and 70% (g, h), respectively. As these panels show, we observe different self-pulsing conditions around the attractor when increasing (e, g) and decreasing (f, h) the gain, even for the same cavity parameters, thus attesting to a variety of multistable regimes.

Besides demonstrating the intrinsic control of the pulse properties dynamics, in our experiments we have found several types of multistable dynamics, as hinted by Figs. 4(a, b). Such a behavior, characteristic of complex nonlinear systems, is investigated in more detail across our tunable parameter space in Fig. 5. As can be seen in Figs. 5(a, b), a greater amplifier gain does not necessarily correspond to an increase in the lasing intracavity power, which is in fact dependent on the history of the system. More importantly, this multistable behavior can directly impact the operation mode of the laser. This can be observed by comparing the average power shown in Figs. 5(a, b) with the corresponding peak power maps illustrated in Figs. 5(c, d): even when using the same set of loss and gain parameters, self-pulsing cannot be observed as easily when approaching particular operation regimes with a decreasing gain; in this case, self-pulsing is lost early on, even though the cavity oscillation is featured with a higher average power - see Figs. 5(b, d). Such behaviors have been observed experimentally for a wide variety of conditions, leading to different discretization of the pulsing dynamics around the attractor of the system, as illustrated in the phase portraits reported in Figs. 5(e-h). Physically, and besides small polarization-dependent parasitic effects in our experiments, we attribute this behavior to be predominantly associated with the known bistability of optically-injected resonators [41]. In fact, we have observed similar dynamics in our simulations, further suggesting that all-optical control of the pulsing operation can be reached, and the intrinsic multistability of the system further exploited.

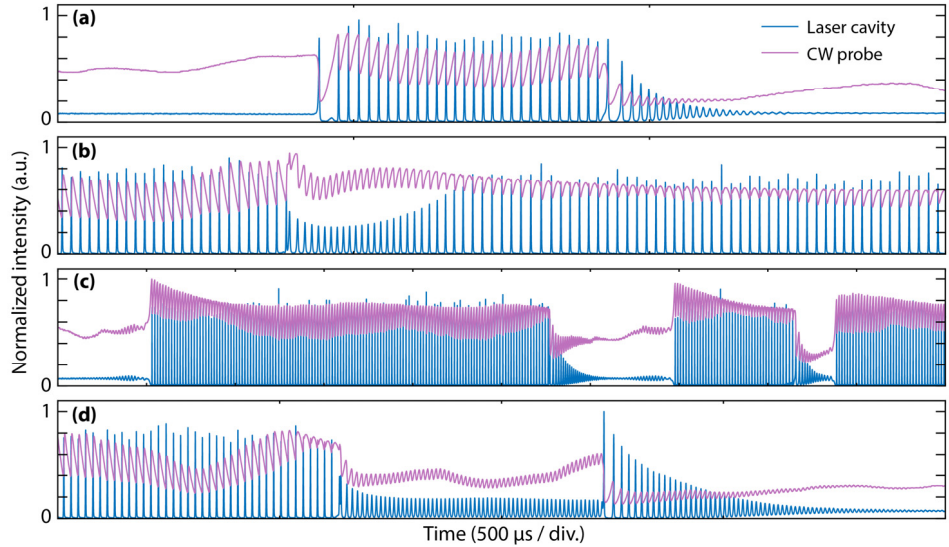

Fig. 6. (a-d) Examples of experimental laser cavity dynamics observed when using a strong counter-propagating CW field (purple line) to both probe and influence the main cavity pulsing properties (blue line). The parameters of the main amplifying cavity are kept constant and a 15 mW CW probe around 1560 nm is coupled to the resonance but not frequency-stabilized. The drift of the CW around this resonance lead to a variety of slow-fast and mutistable dynamics observed in self-pulsing and correlated to the CW field coupling, whose induced thermo-optical effects on the resonator properties cannot be neglected anymore.

To assess this possibility, we have conducted a proof of principle experiment using the same setup as the one shown in Fig. 1(a) but inserting a high-power counter-propagating CW probe in the resonator. Differently from the previous configuration, the effect of the CW probe on the thermal and optical state of the resonator can no longer be considered negligible and it is expected to have a significant impact on the properties and dynamics of the main laser cavity. In particular, we used a 15 mW CW field to seed the resonance at around 1560 nm, while the main laser cavity was oscillating within the resonance located at 1545 nm and operating in a self-pulsing regime with an average power of 30 mW. Although the CW field possessed a power with the same order of magnitude as the lasing cavity (and thus expected to have a similar impact in terms of thermo-optical effects), we observed that similar self-pulsing properties could be maintained or recovered (via e.g. polarization adjustments) over a large range of parameters when the CW coupling properties (i.e. frequency detuning and power) were experimentally stabilized. This can be understood by the fact that the thermo-optical shift induced by the CW field might, in this case, be constant and only lead to an additional offset detuning in the laser-resonator coupling and operation (as long as the interaction between the CW and self-pulsing thermo-optical effects are small enough to avoid crossing any stability boundaries in the dynamical system – i.e. the typical cases shown in Figs. 3(c, d)). However, when the frequency stabilization of the CW probe is turned off, the CW laser can drift in and out of the resonance at the typical microsecond timescales at play in our system self-pulsing dynamics. Examples of such behaviors are shown in Figure 6, where one can see that the fast modification of the CW probe coupling (purple line) drastically influences the main laser cavity self-pulsing dynamics (blue line). Depending on the conditions, variations on the CW detuning and associated coupled power can initiate self-pulsing (Fig. 6(a)), exhibit transient behavior due to the passage from blue to red frequency detuning from the resonance (Fig. 6(b)), display slow-fast dynamics associated with complex and threshold-like coupling behavior (Fig. 6(c)), or even yield frequency doubling of the self-pulsing operation (Fig. 6(d)).

These results clearly underline the capability of an external field to drastically influence the resonator state via the thermo-optical effect, and thus impact the pulsing dynamics of the system in a non-trivial manner. In this framework, we expect this architecture, displaying multistability with widely adjustable pulsing properties, to provide an excellent testbed for the experimental study of complexity in nonlinear optical systems, with timescales easily handled by conventional electronic systems. Additionally, our experimental observations using a CW probe demonstrated the ability to transfer dynamics between multiple wavelengths, thus opening up potential pathways for the wavelength-independent control and synchronization of pulsing in multiple systems.

## 4. Conclusion

We have experimentally demonstrated sustained self-pulsing driven by the complex interaction between the thermo-optical effect in a microresonator and the gain of an external amplifying fiber loop. Numerical simulations provide direct insight on the dynamics of the thermo-optic frequency shift in the resonator and are in good qualitative agreement with experimental measurements made with an auxiliary probe laser coupled to the resonator. Our system exhibits a multitude of interesting multistable dynamics and we have experimentally investigated the effect of a probe field with non-negligible power, with results hinting towards possible applications in complex lasing and nonlinear dynamics, with modes that can be linked by thermally-dependent and adjustable effects.

**Funding**

We acknowledge the support of the EPSRC, Industrial Innovation Fellowship Programme, under Grant EP/S001018/1, from INNOVATE UK, project 'IOTA' grant agreement EP/R043566/1 and from the University of Sussex RDF programme. This project has received funding from the European Research Council (ERC) under the European Union's Horizon 2020 research and innovation programme Grant agreements n° 725046 and 756966. Maxwell Rowley acknowledges the support of the EPSRC through the studentship EP/N509784/1. Benjamin Wetzel and Juan Sebastian Totero Gongora acknowledge funding from the Helena Normanton Fellowship of the University of Sussex. Leonardo Del Bino acknowledges funding from EPSRC through the CDT for Applied Photonics. Jonathan Silver acknowledges funding through a Royal Academy of Engineering fellowship.